# Large Language Models for Departmental Expert Review Quality Scores

Liv Langfeldt, Dag W. Aksnes, Henrik Karlstrøm, Nordic Institute for Studies in Innovation, Research and Education (NIFU), N-0608 Oslo, Norway.
Mike Thelwall, School of Information, Journalism and Communication, University of Sheffield, UK.

**Abstract**

Presumably, peer reviewers and Large Language Models (LLMs) do very different things when asked to assess research. Still, recent evidence has shown that LLMs have a moderate ability to predict quality scores of published academic journal articles. One untested potential application of LLMs is for internal departmental review, which may be used to support appointment and promotion decisions or to select outputs for national assessments. This study assesses for the first time the extent to which (1) LLM quality scores align with internal departmental quality ratings and (2) LLM reports differ from expert reports. Using a private dataset of 58 published journal articles from the School of Information at the University of Sheffield, together with internal departmental quality ratings and reports, ChatGPT-4o, ChatGPT-4o mini, and Gemini 2.0 Flash scores correlate positively and moderately with internal departmental ratings, whether the input is just title/abstract or the full text. Whilst departmental reviews tended to be more specific and showing field-level knowledge, ChatGPT reports tended to be standardised, more general, repetitive, and with unsolicited suggestions for improvement. The results therefore (a) confirm the ability of LLMs to guess the quality scores of published academic research moderately well, (b) confirm that this ability is a guess rather than an evaluation (because it can be made based on title/abstract alone), (c) extend this ability to internal departmental expert review, and (d) show that LLM reports are less insightful than human expert reports for published academic journal articles.

## 1   Introduction

Academic promotion and appointments processes and research assessment systems rely heavily on expert judgement. For research-related personnel decisions, departmental or external experts may review the quality of candidates' work. Perhaps most formally, in the UK, the Research Excellence Framework (REF) evaluates research outputs primarily in terms of originality, significance and rigour, using a four-point scale anchored by qualitative descriptors (REF2021, 2019). All articles in this process are formally evaluated at least twice: once by departments deciding which outputs to submit and once by the external assessors giving the final score. In parallel with longstanding debates about the costs, burdens, and variability of peer review (Aczel et al., 2021), particularly at the scale of the REF evaluations (Wilsdon et al., 2015), the rapid diffusion of large language models (LLMs) has created new interest in whether automated text-based assessment can support or partially substitute for human evaluation. This article focuses primarily on internal departmental assessment, although it is informed by, and has implications for, other types of post-publication quality evaluation.

Early uses of AI in research evaluation have tended to focus on efficiency gains rather than replacement. For example, one funder has trialled AI-assisted prescreening to identify proposals unlikely to be funded, with humans retaining decision authority, raising practical as well as ethical and legal considerations (Carbonell Cortés et al., 2024). In publishing, LLMs have been explored as tools for generating feedback on manuscripts, with evidence that model-generated comments can overlap with human reviewer points at levels comparable to the overlap between two human reviewers in some contexts (Liang et al., 2024). These developments motivate a closer empirical examination of what LLM-based "reviews" capture when asked to operate under a formal assessment rubric.

A key uncertainty is whether LLMs are predicting quality in a meaningful sense, or whether they are reproducing generic evaluative language and superficial correlates of perceived "good science", for example by noting the presence or absence of conventional methodological sections or research limitation discussions. Prior studies suggest that score predictions from appropriate prompts and sufficiently powerful LLMs correlate positively with human assessments in all fields, but that correlations vary substantially by field and input type, and often improve when multiple model outputs are averaged (Thelwall & Yaghi, 2025b; Thelwall & Yang, 2025). Nevertheless, the strongest positive correlations have been obtained when the LLMs were fed with article titles and abstracts rather than their full texts, so the scores are not "evaluations" in the conventional sense but only pattern-based guesses (Thelwall & Yang, 2025).

In the context of the above discussion, three key issues have not yet been addressed and are the target of the current article. First, all previous evaluations of LLMs have used pre-publication peer review scores or ratings (Thelwall & Yang, 2025), external expert review scores (Thelwall & Yang, 2025; Wu et al., 2025), or scores derived from the research team (Thelwall, 2025a). No study has yet used internal departmental review scores, despite this is common use in the UK and countries REF-like national research evaluation systems. Second, although there have been extensive comparisons of LLM comments and pre-publication peer review comments (e.g., Zhou et al., 2024), no study has compared LLM comments with post-publication expert review comments. This is an important omission because pre-publication comments focus on what can be changed in a paper and whether it is publishable, whereas post-publication expert review comments might be expected to focus instead on overall considerations of quality. Insights into the LLM-expert differences may shed light on how LLMs are able to guess scores without evaluating articles. Third, all previous post-publication studies have used quality scores generated by the research team (Thelwall, 2025a) or quality proxies (Thelwall & Yang, 2025; Wu et al., 2025), so an evaluation with direct external quality scores will help to confirm the general LLM quality scoring ability.

This study addresses the above three issues by examining how LLMs score and review journal articles when instructed using REF-like criteria. It compares model-generated scores and review texts with REF-linked assessments produced by field specialists from a departmental evaluation process (from the University of Sheffield Information School) for information science journal articles from the same department. The overarching aim is not only to quantify predictive capacity in this new context, but also to better understand what kinds of evaluative signals LLMs appear to use when generating REF-like judgements. Specifically, the following research questions will be addressed.

- RQ1: Do LLM-generated peer-review scores correlate positively with departmental expert review scores?

- RQ2: How does the strength of the answer to the above question vary between full text and title/abstract inputs and between LLMs?
- RQ3: How do post-publication quality evaluation reports compare between expert and LLM reviewers?

# 2 Background

## 2.1 Research quality scoring with LLMs

There is now a substantial amount of research into potential uses of LLMs for research assessment, both before and after publication. In pre-publication settings, LLMs have been used like reviewers by producing narrative reviews, structured critique, or acceptance recommendations, sometimes benchmarking outputs against editorial decisions or reviewer scores (Chauhan & Currie, 2024). They have been shown to be moderately accurate at predicting scores or decisions (Zhou et al., 2024), at identifying the same strengths and weaknesses as human reviewers (Du et al., 2024), and in giving useful feedback to authors or editors (Lu et al. 2024; Saad et al., 2024). One consistent finding is that LLMs can often generate plausible review language and identify common methodological issues, but may be less reliable in assessing novelty, importance, or field-specific contributions (e.g., Du et al., 2024).

Large-scale evidence suggests that LLM feedback can resemble human review feedback in non-trivial ways while still exhibiting systematic limitations. An analysis of thousands of papers across multiple venues found that the overlap between GPT-4-generated feedback and human reviewer points can be comparable to the overlap between two human reviewers in those venues, while also documenting limitations and variation across contexts (Liang et al., 2024). These results support the idea that LLMs can contribute useful critique, especially for identifying standard issues that are visible from the manuscript text alone. At the same time, they do not establish that model outputs can replicate expert evaluation as practiced in selective peer review or national assessment exercises.

A related strand focuses more directly on predicting post-publication expert quality scores rather than generating feedback. For REF-like evaluations, the LLMs ChatGPT, Gemini, and Gemma3 can produce weak-to-moderate correlations with external or proxy measures of research quality in some contexts (Thelwall, 2025ade). Averaging scores from multiple runs consistently improves the results, but their value is similar whether the LLM is fed with the full text or just the title and abstract (Thelwall & Yaghi, 2025b; Thelwall, 2024; Thelwall, 2025ae). The limited evidence so far suggests that Gemini works better with full texts, but ChatGPT works better with just article titles and abstracts. A consistent issue is that score distributions are compressed (often clustering around mid-to-high values), and are typically higher than human scores. Because of this, individual score predictions are usually meaningless, but they become useful when used to rank articles in a set. For example, a LLM score prediction of 3.05 is almost useless until it is translated into a percentile rank (e.g., in the top 10% of LLM scores from the set of articles analysed). Another issue is the potential for LLM biases, such as for or against certain methodologies or topics (Nunkoo & Thelwall, 2025), and against older research and papers with short abstracts (Thelwall & Kurt, 2025).

From a wider perspective, LLMs also have some capacity to support broader research evaluation tasks, such as assessing narrative claims about the impact of academic research (Kousha & Thelwall, 2025) or the robustness of departmental research environments (Kousha et al., 2025). In these settings, an additional concern is that models may incorporate reputational cues, field conventions, or training-data artefacts that correlate with perceived prestige rather than intrinsic quality. This seems less likely for article evaluation because LLMs do not seem to embed knowledge about individual academic articles, such as the journal they were published in, who wrote them, and when (Thelwall, 2025e).

## 2.2 Human and LLM reviewers

The (post-publication) expert and (pre-publication) peer review of academic outputs like journal articles and conference papers are usually based on field-specific expertise. Reviewers are usually selected for their expertise in the research field of the paper, either narrowly (e.g., for topic expertise or methods expertise in journal review) or general field knowledge, as can be the case in departmental or national expert review. It seems likely that reviewer expertise is most aligned to the evaluated article when the editor/manager has either a large pool of potential reviewers, or free choice (e.g., for journal review) rather than a limited set (e.g., departmental reviews, the UK REF). In either case, reviewers may also use some cross-field expertise about general scientific standards, stringency or clarity, while their level of expertise in the specific research topic and methods of the paper may vary greatly. Still, the key meaning of 'scholarly peer review' is that research is assessed by *peers*, in the sense that reviewers and reviewed have the same or similar competencies.

Considering that modern science is highly specialised, few are deemed qualified to review work within a specific academic field (Chubin & Hackett 1990). The required level of expertise may vary somewhat depending on the aspects of research that are assessed. For example, assessments of scholarly value/significance and originality may demand in-depth knowledge of the state of the art in a specific field, while assessments of methodological soundness/rigour may require methodological expertise that a wider group of reviewers have, unless it is a specialized or uncommon method. Hence, peer/expert review implies that reviewers (are expected to) search for answers within the knowledge pool of their academic discipline and field of research, and in their knowledge communities' concerns to advance scientific knowledge (Langfeldt et al. 2020).

In contrast to expert/peer reviewers, LLMs are trained on enormous corpora of texts rather than primarily on specific fields of research or knowledge communities. Generative LLMs (the type analysed here) also have the ability to generate reliably grammatically correct and meaningful text (Radford et a., 2019), albeit with a risk of hallucinating (Sriramanan et al., 2024). Despite their vast training corpora, LLMs also have the facility to select from their current "knowledge" (Vaswani et al., 2017) so that they can respond appropriately and seamlessly to requests for cake recipes, high energy physics reviews or arts evaluations. Thus, when given an academic work and asked to assess it, they may draw upon the reviewed text as well as reviews and related information that they have met before and consider relevant. The output is then likely to reflect an unknown combination of patterns extracted from apparently related texts. Since LLMs learn using a probabilistic approach and remember knowledge better when it is repeated, they seem likely to recall and replicate more general review points (e.g., "The literature review is inadequate."), much better than more specific critiques (e.g., "The Old Norse discussion should mention ambiguous agency in middle voice verbs.")

because the latter would be individually rarer. Individual comparisons are needed to be sure of the differences, however, because of the complexity of LLMs, including the processes enabling them to respond appropriately to human requests (Ouyang et al., 2022).

## 3 Methods

The first part of the research design was to obtain a set of journal articles with associated departmental reviews and scores. The second part was to submit them to a range of LLMs with different input types (i.e., full text, title/abstract) and compare the results with human expert evaluations of the same articles. Since previous studies have established that the results are more useful if scores from multiple trials are averaged, each article was submitted 30 times in each configuration and the scores averaged.

The third part involved computing inter and intra rater agreement rates, and evaluating text similarity. Lastly, two of the authors read and qualitatively compared a selection of the human and AI review reports.

### 3.1 Data

The data for this paper comprised 58 journal articles from the University of Sheffield (former) Information School in the UK, together with private internal quality ranks from anonymous reviewers within the school, averaged and adjudicated by a team of senior researchers from the school. For each article there was also a summary review written by the school REF head from the original individual reviews.

As part of REF preparations, Information School staff are periodically asked to submit their best work for internal departmental review. The scores given internally, are private and only normally known to the reviewers, the article authors and the school's REF management team. They are used to help select work for the eventual REF submission from the school and not for promotion or appraisal. Two reviewers per output are chosen from the school by the school's REF management team, who conduct single blind review and recommend a REF score. The school's REF management team then agree a final rating (not a score) and combine the feedback of the two reviewers into a single report that is sent to the submitting author with the agreed rating. The main differences between these ratings and external REF scores are that (a) the school staff are likely to be less senior, on average, than REF assessors and (b) staff are not given direct score predictions but an informal indication of score likelihoods, which can be translated into four different score levels.

Since the reports and score levels are private, permission was needed from the school's REF management team and the individual academics to access them. After ethical review, this was obtained by circulating a permission request accompanied by a web form explaining the data uses. Staff completing the form were asked to approve the use of their submitted outputs for the study. The school's REF management team supplied the identities of the outputs, review reports and score levels for the staff that gave permission. The journal articles were used as the raw data for this study. The grey literature outputs were not used because they are substantially different types of documents in the sense of being much longer than journal articles and formatted for a different audience (practitioners). The book chapters could have been used but they lacked abstracts and since the abstract has been critical in previous studies, it did not seem reasonable to create them for this exercise.

In contrast, some of the conference papers were equivalent to journal articles. In all cases these were full papers in the computing field, where conference papers are often more important than journal articles, and so they are the same type of document. Nevertheless, in contrast to the journal articles, their full text contained numerous formulae and mathematical symbols that were impossible to describe in a standard plain text file. The source files of these PDF files were LaTeX document description files. These could have been requested from the authors, but this would have meant comparing plain text files with LaTex document descriptions, making an unequal comparison. Thus, to avoid the possibility of skewing the results because of this, all conference papers were excluded and only journal articles retained. One article was an unusual short-form two-part document where the journal published a two-page summary and linked to a self-published extended version. This was excluded for being a different type of document. These sampling decisions were all made before submitting any to LLMs to avoid biasing the decisions from the results.

For convenience, the school score levels were allocated on a four-point REF-like scale that for convenience is labelled 1*, 2*, 3*, or 4*, but the actual labels used have been redacted for policy reasons.

Abstracts and PDFs were obtained online for the papers submitted, along with the copyright status of each paper. The PDFs were edited to remove all journal and author identifying information to reduce the chance that LLMs could leverage publisher or author information in making their scores. This process included removing page numbers and running headers and may have also helped the LLMs to identify the core text in the papers separately from publishing-related information. A copyright check was performed on each paper to identify ownership and reuse conditions. The owners were either the author(s) or the publishers. In the former case we had permission for reuse. In the latter case the copyright statement was always a version of CC BY, permitting re-use with attribution. The latter does not clearly allow use in public LLMs that learn from their inputs because they generate material without attribution. Thus, only LLM interfaces with guarantees not to learn from inputs were used. The following datasets were derived.

- 58 journal article titles and abstracts. Under UK law these can be freely republished and so can be used for machine learning.
- 58 journal article PDFs, with metadata redacted as described above. These include PDFs with author copyright ownership and PDFs with journal ownership with copyright conditions allowing republishing without attribution. In the UK there is a research copyright exemption for text processing that includes lawfully accessed copyright content.
- 58 full text documents containing all the text in the PDFs except the figures. The copyright situation is as above.
- 58 truncated text documents containing all the text in the PDFs except the figures and references. The copyright situation is as above.

Recall that previous studies with LLMs evaluating published journal articles either used published articles with public indirect score information (Thelwall, 2025b) or private scores from a single author (Thelwall, 2024). Thus, the current study is the first with these prompts to use private scores and articles from different authors (one of the scorers and article authors was the same, but none of the articles are the same as any prior test).

### 3.2 Prompts

Most generative LLMs can be configured for a particular task by submitting system prompts (instructions) before the user prompt. The system instructions define the task and can also give style guidelines for the output. The same system prompt as use in previous similar studies was reused. Specifically, the prompt is the REF2021 instructions for evaluators in REF Main Panel D (arts, humanities and some social sciences) (REF2021, 2019), with small wording changes to adapt it to the ChatGPT instruction style. The exact text of the prompt is in an online appendix (Thelwall & Yaghi, 2025a), and the instructions centre on rigour, originality and significance, which are usually believed to be core components of research quality (Langfeldt et al., 2020).

The user prompt for each request was "Score this article" followed by the article text. A previous study has tested variations of this prompt and variations of parameter settings in ChatGPT, but none improved the results (Thelwall, 2025a) so only the standard REF prompt and the default ChatGPT settings (temperature=1, top_p=1) were used. The default Gemini settings were also used.

### 3.3 LLMs

Previous attempts to predict quality scores for published journal articles with LLMs have mainly used ChatGPT alone but one has used Google Gemini Flash 1.5. Some have used downloaded smaller quantised versions of LLMs (cut down versions with lower accuracy), and they have also been used to support pre-publication submission assessment, with positive results (Du et al., 2024). This article uses the full ChatGPT-4o (gpt-4o-2024-11-20), the cheaper ChatGPT-4o mini (gpt-4o-mini-2024-07-18), and Gemini 2.0 Flash, Google's competitor to ChatGPT-4o mini at the time of data processing. Only Gemini was tested for PDFs because the ChatGPT API does not allow automatic submission of documents. All queries were submitted in March 2025.

### 3.4 Computed metrics

The association between AI-generated scores and human ratings was assessed using both Pearson's product–moment correlation coefficient (r) and Spearman's rank-order correlation coefficient ($\rho$). Pearson's r was included to assess the degree of linear relationship between the scoring systems, although one is ranks, and Spearman's $\rho$ to account for the ordinal nature of the human rating scale and potential non-linear but monotonic associations. Spearman correlation was the primary measure to compare the LLM score predictions with the human scores, reflecting the likely use case for LLM scores is to rank a set of publications.

Previous studies have found that LLM scores tend to cluster around 3*, often tightly (e.g., Thelwall, 2024). Whilst it would be possible to create a lookup table to convert these values into the full range 1*, 2*, 3* or 4* this would be relatively arbitrary and lose information. In contrast, ranks are unaffected by this clustering. They would also closely match the way in which citation data is used to support some UoAs in REF2021, which was by reporting the citation percentile for articles. The ranks are also the most important factor for REF departmental evaluations since each department submits its top n publications for national evaluation, where n is determined by a formula based on the number of research active staff on the census date.

For measuring text similarity between model outputs, LLM reports were represented as vector space text embeddings, which has been shown to better capture semantic meaning than traditional text similarity measures such as the Jaccard or cosine similarity (Khosla et al. 2024). For each review, the vector embedding was retrieved using a dedicated embedding model, and the cosine similarity of the vectors was computed for each pair of reviews, both within and across models. The similarity score was averaged across each batch of thirty scores, producing three intra-model similarity scores for each article and a pairwise similarity score between each model.

### 3.5 Text analyses

With the aim to understand how LLMs predict peer review scores, the strategy was to examine LLM and human review text of papers with *low* AI and human score correlation and compare with the reviews of papers with *high* AI and human score correlation. For this part, only three datasets were used: the ChatGPT-4o reports based on full text and based on title/abstract; and Gemini 2.0 Flash reports based on the full text. ChatGPT-4o had the strongest correlations with expert review scores in previous studies, which is why it was selected. Previous studies had also found the title/abstract input to be the best, so it was a logical choice. Full text input is also a logical choice to match the reviewers, since this is closer to what they would see. Although the reviewers would read the PDF files, these can't be processed by the ChatGPT-4o API used here. The Gemini 2.0 Flash reports based on full text were included to parallel the ChatGPT-4o full text reports so that there would be a chance to compare the three sources of analysis of full text (human, ChatGPT, Gemini). Although the PDF input could have been used for Gemini for this, using the same input as ChatGPT should help to highlight potential differences between models.

Articles for the qualitative analysis were selected in a 2-step procedure. First, based on the human score and AI inter-rater agreement for each paper, we selected the articles with the highest and lowest human-AI score match. The inter-rater agreement was calculated using Krippendorff's alpha. For each human score category (REF score 1, 2, 3 and 4) we selected the papers with the lowest and highest AI inter-rater agreement. This resulted in a sample of 15 articles (of the total of 58 articles in the quantitative analyses), with score distribution as displayed in Table A1 in the Appendix. In a second step, we selected the first (of the 30) appearing review text in each model that gave an overall score similar to the model's average rating. This resulted in a sample of 60 reviews (15 human and 45 AI reviews) for the text analysis.

Two of the authors read the human and AI review reports and the abstracts/papers. We started with a test round in which we coded the same 10 human reviews separately and then compared the results. This initials coding scheme coded review comments as positive, negative, both positive and negative or indecisive/not commented for each of the review criteria (Originality, Significance, Rigour), whether these comments were detailed or general, and whether criteria other than the three REF criteria were commented. Reviewing the results of this initial exercise, we concluded that the coding was not helpful for understanding differences between the reviews and that a more explorative analysis was needed.

The analysis in this paper is based on consecutive reading of the one human and three AI reviews of each article (including the article abstract and checking content of the article when relevant) while taking notes of similarities and differences between the reviews. We then compared results for cases where the LLMs were able, and cases where they were not able, to predict the human scores, as well as cases where the models rated lower than the human score and those where the models rated higher than the human score. The analysis was aimed at finding patterns that could help understand differences and similarities.

## 4 Results

The quantitative evidence is reported first, followed by the qualitative findings.

### 4.1 Correlations between LLM and departmental expert scores (RQ1, RQ2)

Overall, the LLMs assigned high mean scores, with all ChatGPT averages being above 3 and all Gemini averages being at least 2.73 (Figure 1). For each model, the lowest scores occurred for the title/abstract inputs. As suggested above, this might be due to articles being penalised for omitting information that was in the full text. Recall that the human score levels are nominal and not directly comparable.

In terms of the variability of scores (i.e., average of 30 scores per paper) between articles for a single model and input, for ChatGPT there was a range from below 3 to 4 or near 4, although with much lower values for title/abstract input (Figure 1). For full or truncated text input, the lowest score given to any article by any ChatGPT was 2.89, whereas for title/abstract input it was much lower at 2.13.

Gemini had a very different variability pattern. With the partial exception of title/abstract inputs, it had a very strong preference for the score 3, rarely giving less and almost never giving more. For example, with PDF input there were 58*30=1740 queries, 1657 of which (95.2%) returned 3. Thus, Gemini is highly confident that almost all 58 articles are worth 3 (i.e., almost always predicting this in individual reports), whereas ChatGPT is much more willing to consider other scores, from 2 to 4 (i.e., more frequently mentioning them in reports for individual articles).

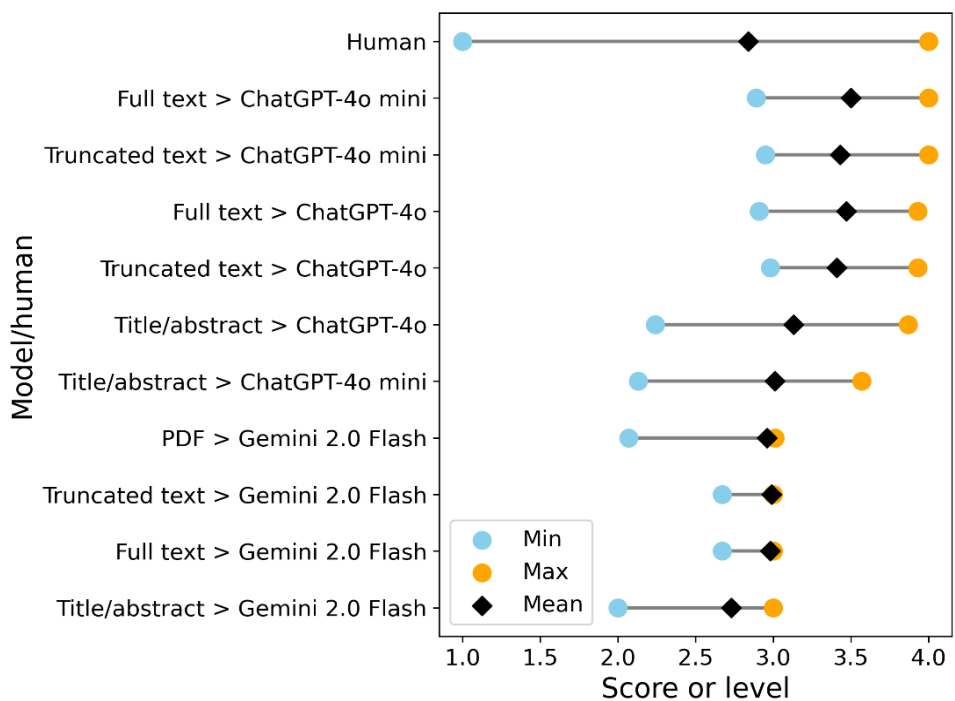

Figure 1. Minimum, maximum and mean scores for the different LLM and inputs, and the same for human (i.e., internal departmental) nominal score levels (n=58 articles). The LLM figures were calculated *after* averaging 30 iterations.

Irrespective of the LLM, in all cases there is a tendency for articles with a higher human score level to get a higher LLM score (Figure 2). There is a big overlap between LLM scores at different levels, however, with no LLM score range mapping onto a single human score level. Thus, there is no case where the human score can be deduced with certainty from the LLM score.

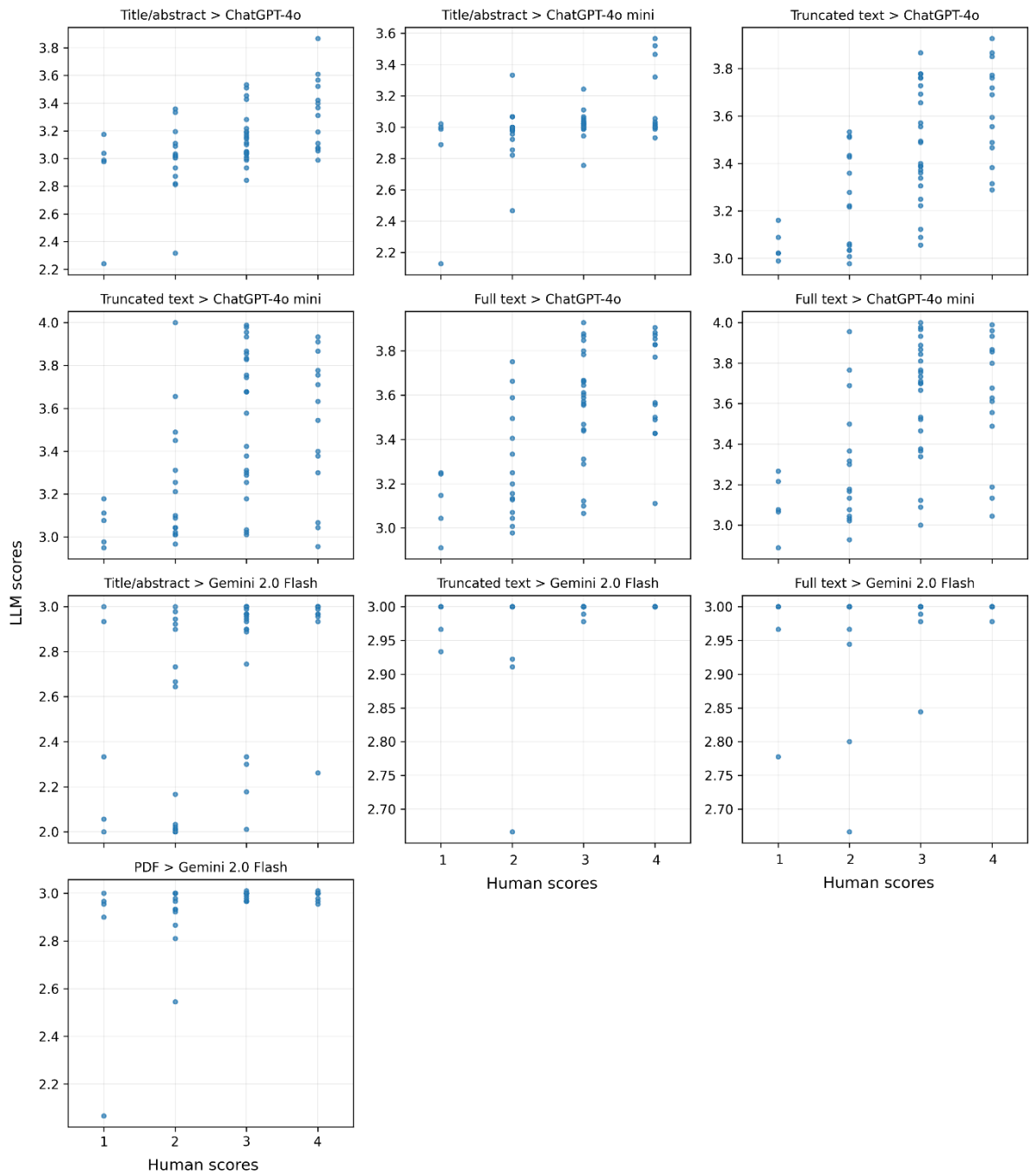

Figure 2: Departmental score levels against LLM score predictions for 58 information science journal articles.

In all cases, the Spearman correlation between the LLM scores and human score levels for the 58 information science articles was positive, with the value being statistically significantly different from 0 except for one marginal case (Figure 3). Unfortunately, the 95% confidence intervals are too wide to get statistical evidence of which LLM or input is best, but generally ChatGPT-4o performs best (the highest three correlations) and none of the four inputs seem to be consistently the best or worst. The only document format containing all information necessary for an evaluation (text, figures and references), PDF format, has the third lowest correlation in Figure 3, however, which tends to confirm that LLM scores are guesses rather than judgments. The results are similar for Pearson correlations (Figure 3).

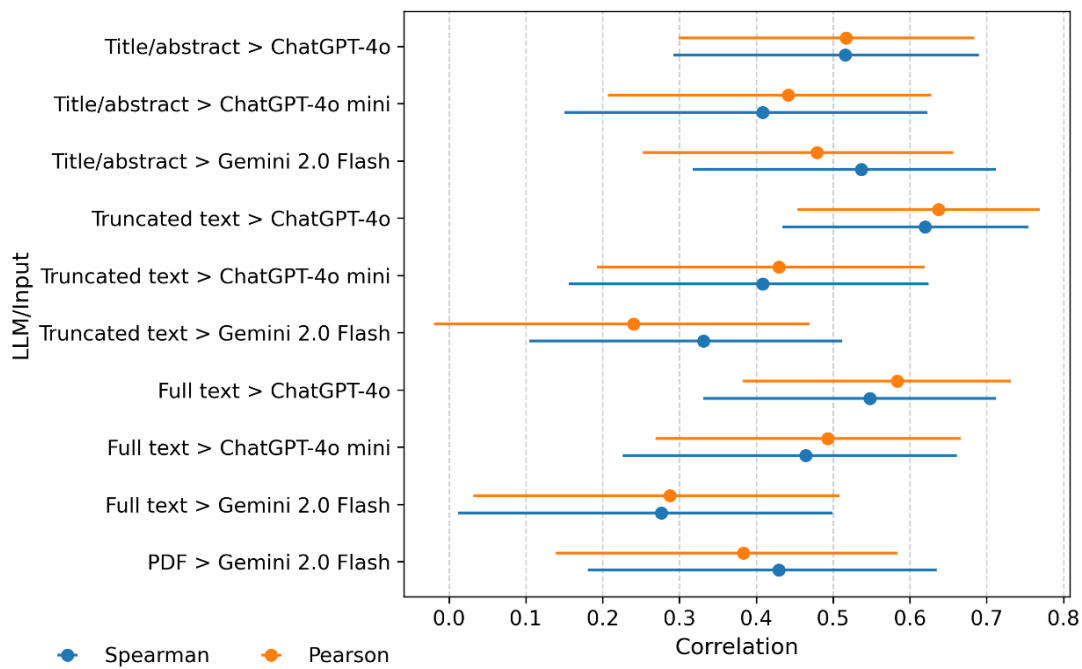

Figure 3. Spearman and Pearson correlations between LLM score predictions (after averaging 30 iterations) and human score levels by LLM and input (n=58 articles). Bootstrapped 95% confidence intervals are shown.

A heatmap was drawn to illustrate how the scores are similar between LLMs (Figure 4). For ChatGPT, the correlations between versions and inputs tend to be higher than the correlations with the human scores, but the Gemini correlations tend to be universally low, except for title/abstract inputs.

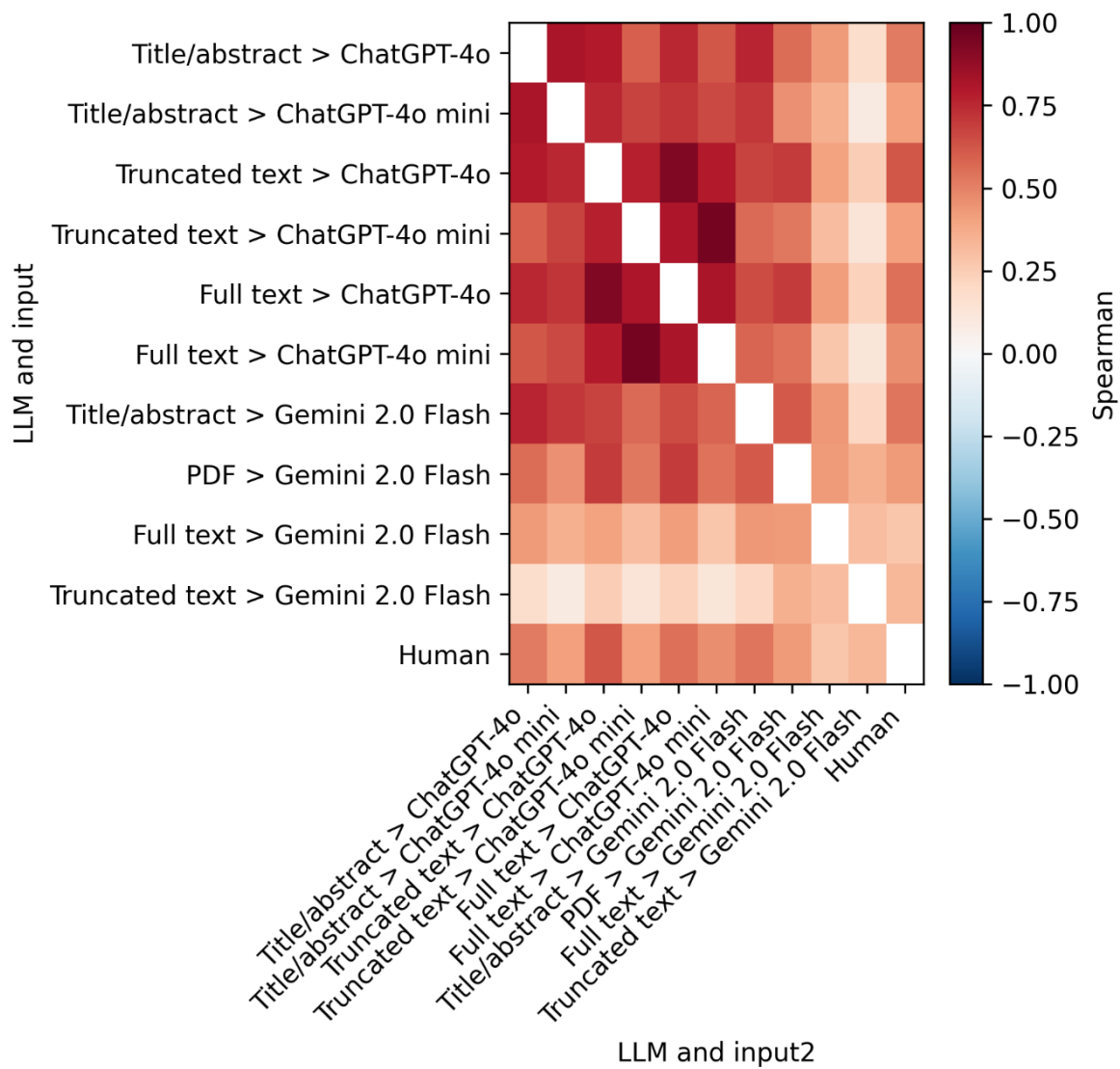

Figure 4. A heatmap of pairwise Spearman correlations between LLM score predictions (after averaging 30 iterations) and human score levels by LLM and input (n=58 articles).

### 4.2 *Quantitative evidence of text similarity in LLM-model reports (RQ3)*

The mean text similarity between the 30 review reports generated for each paper and model was very high. For all 58 papers, intra-model similarity had only a narrow range of variation. For example, similarity scores ranged from 0.92 to 0.97 for ChatGPT-4o with full text inputs, from 0.91 to 0.97 for ChatGPT-4o with title/abstract inputs, and from 0.91 to 0.96 for Gemini 2.0 Flash with full text inputs. Thus, in all cases when a LLM was fed with the same output repeatedly, its reports tended to be very similar, despite never being identical and sometimes reporting different scores.

Cross-model text similarity was also high, particularly between the two ChatGPT variants, and even for different inputs for the same paper. The mean similarity between ChatGPT-4o with title/abstract inputs and ChatGPT-4o with full text inputs ranged from 0.89 to 0.97. Slightly lower similarity scores occurred when either ChatGPT variant was compared with Gemini: 0.85–0.94 for ChatGPT-Abstract versus Gemini and 0.85–0.93 for ChatGPT-Full versus Gemini. Across all model pairings, the median similarity values were located close to the midpoint of each respective range. Overall, these results show that both intra-model and cross-model variations in review text are minimal. The high similarity scores suggest that the models tend to produce highly repetitive and standardized review reports, with limited diversity in wording and content.

Inspection of the LLM review reports revealed several interesting patterns. In general, these reports were substantially longer than the human-written departmental review summaries. The texts could be characterised by *wordiness*, using many words to explain a single point and often repeating it using different formulations. For example, one 180-word abstract was assessed by an 800-word ChatGPT review.

## 4.3 *Qualitative comparison of LLM and human reports (RQ3)*

A qualitative examination of the review reports explains some aspects of the patterns identified above. The LLM-generated reviews tended to be more positive in tone than the human review summaries. While it is difficult to determine precisely why they differ in correspondence with human scores, an important difference appears to lie in how the models use the rating scale. ChatGPT-4o reviews based on titles/abstracts, which are generated from limited information, tend to be more restrictive in its scoring. In contrast, ChatGPT-4o assessments based on full text assign higher scores. As mentioned above, Gemini, by comparison, produces scores that are highly compressed around the midpoint of the scale: nearly all articles are rated close to 3. Gemini's review texts are also generally shorter, less repetitive, and less positive in tone, resembling human reviews more closely in style.

Overall, the AI-generated texts appeared coherent and meaningful. The models seem to be well trained on similar evaluative texts, particularly with respect to research quality, standard review criteria, and the strengths and weaknesses of different scientific methods. Whilst the system prompt was lengthy (one and a half pages), it did not contain information about scientific methods so this would have had to be elicited by the LLMs from their existing knowledge. In their assessments, the LLMs primarily focused on broad characteristics of research quality, such as generalizability, methodological novelty, theoretical contribution, and reproducibility. Thus, the limitations commonly reported accordingly included issues such as low response rates and limited generalizability. In the reports, evaluative terms from the system instructions (e.g., "world leading") were often used without much justification or explanation of what these concepts would mean in practice.

Unexpectedly, when assessing abstracts only, ChatGPT did not seem to adjust for the limited information available, and limitations were sometimes reported in review reports that were incorrect because the points had been addressed in the full texts of the papers.  For example, one review stated: "However, the study would benefit from more detailed discussions about the sample size, participant characteristics, data coding methods, and validation of findings to reinforce rigour" (case 25). This comment is misleading, as the full article provides descriptions of these elements.

Another important difference is that the LLMs primarily emphasize general characteristics of research quality—such as clarity, generalizability, and methodological soundness—whereas human reviewers tend to be more specific, frequently highlighting concrete issues, contextual details, or domain-specific concerns that are not explicitly mentioned in the AI-generated reviews. This suggests that while LLMs can approximate overall evaluative patterns, they may lack the granularity and contextual sensitivity characteristic of human peer review.

Moreover, the AI-generated texts were more focused on the three review criteria of originality, significance and rigor, while the human reviewers sometimes also commented on other aspects they found relevant, such as pointing out that papers were well-structured and well-written, presented in a way so that lay people would be able to understand, or giving positive comments on papers' extensive bibliography or supplementary material. In other words, while the AI models kept to the instructions and assessed all papers by the same criteria, the human review summaries sometimes included article-specific context in their review.

In addition to delimiting comments to the formal criteria, the AI generated comments often included terms from the criteria and score definitions and so give the impression of strictly following the guidelines. For instance, the term 'world-leading' (defining score 4*) and 'internationally excellent' (defining score 3*) do not appear in the human reviews but frequently occur in the AI-generated texts. In particular, the AI-generated texts often included statements that the work is internationally excellent and/or not world-leading:

> *"The originality, significance, and rigour of the work support its contribution to the field, although some limitations prevent it from reaching the highest level of world-leading quality" (Gemini, case 22).*

> *"The paper demonstrates quality that is internationally excellent in terms of originality, significance and rigour but which falls short of the highest standards of excellence" (Gemini, case 54).*

> *"The paper demonstrates internationally excellent rigour, with a well-structured methodology and an expansive literature review" (ChatGPT-4o full text, case 24).*

> *"the article does not achieve world-leading significance due to its limited demonstration of direct profound influence or transformative potential on current thinking, practices, or policies" (ChatGPT-4o full text, case 15).*

Another observation is that the LLMs frequently commented on the strengths and limitations of research methods, including giving advice on how to mitigate limitations and improve robustness or generality. The human reviews also comment on such issues, but less elaborately and less didactically. Some examples from the AI generated text:

> *"The survey data collection is methodologically sound, with clear steps outlined to ensure anonymity, validity, and reliability; however, the survey's limited sample size and potential non-response bias (acknowledged by the authors) weaken the broader generalizability of the findings" (ChatGPT-4o full text, case 15).*

> *"limitations such as the convenience sampling strategy, lack of control over demographic diversity, and the unknown survey response rate reduce methodological robustness" (ChatGPT-4o full text, case 45). "The data analysis methods are appropriate for the research questions, but the study's limitations, such as the* *convenience sampling strategy and potential bias in the participant sample, should be acknowledged" (Gemini 2.0 Flash full text, case 45).[1]*

*"The methodology section is described in remarkable detail, ensuring full reproducibility. By combining theoretical reasoning with empirical findings and statistical variance controls, the article achieves outstanding scientific integrity. [...] Suggestions to Strengthen the Work Further [...] While the current work comprehensively addresses mid-scale datasets [...], additional analysis [...] on much larger datasets would offer deeper insights into computational efficiency across extreme scales. [The strategy] was shown to work variably across imbalanced datasets. Future refinement of how combinations of [...] post-hoc threshold techniques (e.g., calibration curves) can be integrated into model development workflows could better contextualize its utility" (ChatGPT-4o full text, case 22).*

In the last case, advice for improving rigor was given even when assigning the highest score, which indicates that the models do not always provide a coherent relationship between the score assigned and (all) the review comments. The reason for the more didactic nature of LLMs may be that suggesting improvements to users may be the most common use case for them, so it is reasonable to assume that they tend to default to this, even when not asked to provide help.

Examining the articles where the human score level was low and the LLM score was not, there are multiple cases where the human review points to limited significance of the study and shortcomings in methods, while the AI-generated reviews also point to limitations in significance and methods, but in a more positive tone and without giving low scores. For example, in one case the human review commented that the results could prove useful locally - where the study was carried out - but were not likely to be noteworthy beyond that (case 51). In contrast, the LLMs mentioned that "While some limitations exist regarding sample sizes and generalizability, the study demonstrates international excellence in addressing a crucial issue" (Gemini); moreover "while the work has considerable influence and practical implications, it falls just short of world-leading significance" (ChatGPT-Full). In other words, the LLMs surmised that the work is not world-leading but had difficulties predicting that something is not assessed to be useful outside the site studied.

While the scope of relevance and potential influence are crucial aspects when assessing research contributions, scholars may hold different opinions on such aspects, for example because they perceive or emphasize different values in the research or different needs in science or society. Hence, peer review scores may be generally hard to predict. But why are LLMs reluctant to give low scores?

Part of the answer may be that the AI models were neutral to research topics – in the sense that all topics are presented as important and interesting, while within academic fields there are common – or diverse – views on what are interesting and important topics. Another possible answer is that the LLMs may rely heavily on information in the paper to comment on any limitations in the scope of its relevance or importance, while peer reviewers also rely on the values and interests of their academic field. A harsh judgement of the kind that a study is useless outside the site that was studied may be hard to extract from the paper itself – unless in the unlikely case that this is stated in the paper. Taken together, this implies that low scores may be hard to predict because the AI models were neutral to research topics, and when academic papers presented the importance of their topic and results from a positive point of view (rather than as a critical reviewer), the AI models had more difficulties in predicting low than high scores.

---

[1] Convenience sampling was not commented on in the human review, which simply stated that the sample size was good.

# 5   Discussion

The results are limited to a single country, department and field, and by a sample size that was too small to identify statistically significant differences between LLMs or input size. Moreover, the articles examined are neither a random sample nor generated through a neutral selection process. They are preselected by academics as their best outputs, in line with REF-style submission practices. The results are also limited to three LLMs (a Gemini and two ChatGPTs), and one prompting style, the default model settings. The results cannot therefore be generalised to other countries, fields, and models. They may also improve with fewshot or fine tuning.

## *5.1   Comparison with prior results*

The correlation strengths found in the current paper broadly align with those found for a single author set of information science papers with author scores (Thelwall, 2024) and for a dataset of REF2021 journal articles for the wider category Communication, Cultural and Media Studies, Library and Information Science with proxy quality scores (Thelwall & Yaghi, 2025a). They cast doubt on a previous claim that ChatGPT-4o scores are more informative if the input is just title/abstract rather than full text (Thelwall, 2025a) and the opposite for Gemini (Thelwall, 2025c). The current results therefore cement the previous evidence of the ability of LLMs to predict research quality in this field with a new type of data (internal departmental reviews) and suggest that the difference between title/abstract and full text or PDF input may be minor. There are many possible explanations for the latter point, including the following:

- LLMs can struggle to find relevant information in long documents (Liu et al., 2024), so may focus on the start (abstract) and end (conclusion, similar to abstract, or references, which are irrelevant).
- LLMs do not understand documents and may draw false pattern matching conclusions from the full text, such as criticisms in the literature review.
- Abstracts may summarise the key value of a paper that has already been validated through peer review.

The reluctance to award low scores has been noted before (Thelwall, 2024; Thelwall & Yaghi, 2025a), and the current results therefore tend to confirm it. Possible explanations are discussed above, and an additional possibility is that LLMs try to give high scores as a form of positive feedback to users as part of their more typical use case of helping authors improve their work.

The much wider range of ChatGPT scores compared to those of Gemini is a new observation (e.g., not mentioned in: Thelwall, 2025c). In conjunction with the high proportion (95%) of 3* scores in individual Gemini reports, this suggests that a greater degree of averaging will be needed to effectively leverage Gemini's uncertainty about whether an article might get a higher or lower score; it is this uncertainty that is necessary to separate the articles for ranking.

## 5.2 New findings about LLM reports and their difference with human review reports

No previous study has reported a detailed analysis of LLM review reports, although some have mentioned them. Moreover, none have compared LLM reports with any type of post-publication reports, although there are comparisons of LLM and human pre-publication comments. The finding that LLM comments tend to be positive aligns with similar observations about pre-publication peer review (Zhou et al., 2024), although the current study did not find a greater focus on rigour that has been found for conference paper peer review reports by LLMs (Shin et al., 2025; Zhu et al., 2025). The analysis of text similarities (as embeddings in vector space) finding minimal intra-model and cross-model variations in LLM review texts does not seem to have a parallel in any previous study.

Another new finding is that LLM reviews seem to be more focused on the review criteria, while the human reviews were more specific (echoing the situation for peer review: Zhou et al., 2024) and more varied. Furthermore, they linked scores and comments by stating that the paper deserved a score on a specific level (typically 'internationally excellent' for justifying score 3*) and explained this in a general manner (typically not 'world-leading'/score 4* due to some limitations in originality, significance and/or rigor). Links between score and comments were still partly inconsistent, as limitations were pointed out for the top-rated papers.

The new observation that when only provided with the abstract, ChatGPT sometimes provided flawed comments (pointing to limitations that did not appear in the full text paper), that did not occur for full texts. Still, the correlation with the human score remained about the same. Thus, there may be different limitations in terms of ability to predict human scores when reviewing an abstract and a full paper.

## 5.3 LLM scores for internal departmental review

A natural question arising from this analysis is whether the time-consuming internal departmental peer evaluations when generating REF submissions could be supported by LLMs. These probably take place annually in most UK departments, building for the next REF (REF2029). It seems clear that the LLM results are not good enough to replace the human experts (a similar negative recommendation has been made in most previous LLM studies), but could they support them?

In terms of supporting expert judgements, the models are unable to reliably distinguish between truly outstanding and merely adequate publications. The graph for ChatGPT 4o with truncated text input suggests ways in which the scores can help, however (Figure 2). In the Information School, all articles scored level 4 were provisionally selected for the REF and a subset of articles scored level 3 would also need to be selected to make up the quota. One way to select these articles would be to take those with the highest ChatGPT 4o score, given that the human evaluators could not distinguish between them. In practice, the School would almost certainly prefer the departmental REF team to select the articles, but the LLM scores might provide at least a starting point for this difficult decision.

A second way in which the LLM scores might support the departmental review process is indicating need for additional human review. Limitations in the reviews are likely because information science is highly multidisciplinary and the School has a small pool of potential reviewers. This might involve additional human review of articles with human 4* score level but low LLM scores, and perhaps also check if the 2* articles with the highest LLM scores might be worth 3* or 4*.

The above two suggestions conflict in the sense that the first saves time and the second does the opposite but may improve the results. A decision about whether either is appropriate will need to take into account a department's relative confidence in its human reviewers and the LLM predictions.

# 6   Conclusions

This study compared internal departmental (human) quality ratings and review summaries for 58 information science journal articles with scores and review texts produced by ten LLM configurations. The quantitative results show statistically significant, moderate positive correlations between ChatGPT scores and human scores, whether ChatGPT was given full text or only abstracts, while Gemini had weaker correlations. Differences between models and between input texts were not statistically significant. Taking into account previous studies with related quality analyses, there does not seem to be much difference in LLM predictive power if they are fed with titles/abstracts or full texts.

The text analysis helps to interpret these patterns. Across models, review reports were coherent but highly standardised in wording and content, with minimal variation across repeated runs. The models frequently reproduced language from the guidelines and provided broadly plausible strengths and limitations, but often in a manner that remained general and sometimes weakly connected to the assigned score level. When working from abstracts alone, ChatGPT sometimes inferred limitations that were in fact addressed in the full paper, indicating that the model does not consistently represent uncertainty arising from missing information. At the same time, providing full text tended to increase ChatGPT scores without increasing correlation with human scores, reinforcing the possibility that additional author-framed detail increases positivity without improving alignment with expert judgement.

In terms of support for internal departmental reviews, whilst the correlations are not high enough to suggest the time saving approach of replacing expert reviews with LLMs, the predicted scores may still help decision making. Two possible roles are arbitrating for tied scores and identifying articles with anomalous human scores.

# Appendix

**Table A1: Overview of cases in the study, including human, AI scores, and AI inter-rater agreement. Sorted by human score and AI inter-rater agreement. Qualitative analysis sample in bold.**

| Human rater | GPT full article | | GPT abstract | | Gemini full article | | Diff. human AI av.score | | | AI inter-rater agreement |
|---|---|---|---|---|---|---|---|---|---|---|
| Score | Score | Std. deviation | Score | Std. deviation | Score | Std. deviation | H GPT-full | H GTP-abst | H Gemini | Krippendorff's alpha |
| **1** | **3.03** | **0.18** | **2.93** | **0.25** | **2.90** | **0.31** | **-2.03** | **-1.93** | **-1.90** | **0.02** |
| **1** | **3.10** | **0.31** | **2.97** | **0.18** | **2.97** | **0.18** | **-2.10** | **-1.97** | **-1.97** | **0.04** |
| **1** | **3.17** | **0.38** | **3.00** | **0.00** | **2.97** | **0.18** | **-2.17** | **-2.00** | **-1.97** | **0.09** |
| **2** | **3.00** | **0.26** | **2.90** | **0.31** | **2.93** | **0.25** | **-1.00** | **-0.90** | **-0.93** | **-0.01** |
| **2** | **3.03** | **0.18** | **2.87** | **0.35** | **2.93** | **0.25** | **-1.03** | **-0.87** | **-0.93** | **0.03** |
| 2 | 3.17 | 0.38 | 3.17 | 0.38 | 3.00 | 0.00 | -1.17 | -1.17 | -1.00 | 0.03 |
| 2 | 3.00 | 0.00 | 2.87 | 0.35 | 2.80 | 0.41 | -1.00 | -0.87 | -0.80 | 0.04 |
| 2 | 3.27 | 0.45 | 3.03 | 0.49 | 3.00 | 0.00 | -1.27 | -1.03 | -1.00 | 0.06 |
| 2 | 3.10 | 0.31 | 2.97 | 0.18 | 2.87 | 0.35 | -1.10 | -0.97 | -0.87 | 0.07 |
| 2 | 3.13 | 0.35 | 2.97 | 0.18 | 2.93 | 0.25 | -1.13 | -0.97 | -0.93 | 0.07 |
| 2 | 3.03 | 0.18 | 2.80 | 0.41 | 2.97 | 0.18 | -1.03 | -0.80 | -0.97 | 0.08 |
| 2 | 3.17 | 0.38 | 3.00 | 0.00 | 3.00 | 0.00 | -1.17 | -1.00 | -1.00 | 0.09 |
| 2 | 3.20 | 0.41 | 3.00 | 0.00 | 3.00 | 0.00 | -1.20 | -1.00 | -1.00 | 0.11 |
| 2 | 3.43 | 0.50 | 3.07 | 0.25 | 2.97 | 0.18 | -1.43 | -1.07 | -0.97 | 0.24 |
| 2 | 3.03 | 0.18 | 2.43 | 0.50 | 2.53 | 0.51 | -1.03 | -0.43 | -0.53 | 0.26 |
| 2 | 3.43 | 0.50 | 3.00 | 0.00 | 3.00 | 0.00 | -1.43 | -1.00 | -1.00 | 0.32 |
| 2 | 3.70 | 0.47 | 3.37 | 0.49 | 3.00 | 0.00 | -1.70 | -1.37 | -1.00 | 0.33 |
| 2 | 3.77 | 0.43 | 3.13 | 0.57 | 3.00 | 0.00 | -1.77 | -1.13 | -1.00 | 0.41 |
| **2** | **3.67** | **0.48** | **3.03** | **0.18** | **3.00** | **0.00** | **-1.67** | **-1.03** | **-1.00** | **0.51** |
| **2** | **3.00** | **0.00** | **2.30** | **0.47** | **2.03** | **0.18** | **-1.00** | **-0.30** | **-0.03** | **0.66** |
| **3** | **3.07** | **0.25** | **3.03** | **0.18** | **2.97** | **0.18** | **-0.07** | **-0.03** | **0.03** | **0.01** |
| **3** | **3.10** | **0.31** | **3.03** | **0.18** | **2.97** | **0.18** | **-0.10** | **-0.03** | **0.03** | **0.02** |
| 3 | 3.03 | 0.18 | 2.83 | 0.38 | 3.00 | 0.00 | -0.03 | 0.17 | 0.00 | 0.09 |
| 3 | 3.27 | 0.45 | 3.03 | 0.18 | 3.00 | 0.00 | -0.27 | -0.03 | 0.00 | 0.13 |
| 3 | 3.23 | 0.43 | 3.00 | 0.00 | 3.00 | 0.00 | -0.23 | 0.00 | 0.00 | 0.14 |
| 3 | 3.70 | 0.47 | 3.50 | 0.51 | 3.00 | 0.00 | -0.70 | -0.50 | 0.00 | 0.34 |
| 3 | 3.47 | 0.51 | 3.00 | 0.00 | 2.97 | 0.18 | -0.47 | 0.00 | 0.03 | 0.34 |
| 3 | 3.47 | 0.51 | 3.00 | 0.00 | 2.97 | 0.18 | -0.47 | 0.00 | 0.03 | 0.34 |
| 3 | 3.63 | 0.49 | 3.17 | 0.38 | 3.00 | 0.00 | -0.63 | -0.17 | 0.00 | 0.35 |
| 3 | 3.63 | 0.49 | 3.13 | 0.35 | 3.00 | 0.00 | -0.63 | -0.13 | 0.00 | 0.37 |
| 3 | 3.60 | 0.50 | 2.97 | 0.41 | 3.00 | 0.00 | -0.60 | 0.03 | 0.00 | 0.38 |
| 3 | 3.73 | 0.45 | 3.23 | 0.43 | 3.00 | 0.00 | -0.73 | -0.23 | 0.00 | 0.41 |
| 3 | 3.57 | 0.50 | 2.97 | 0.18 | 2.97 | 0.18 | -0.57 | 0.03 | 0.03 | 0.43 |
| 3 | 3.57 | 0.50 | 3.00 | 0.00 | 3.00 | 0.00 | -0.57 | 0.00 | 0.00 | 0.45 |
| 3 | 3.63 | 0.49 | 3.00 | 0.26 | 3.00 | 0.00 | -0.63 | 0.00 | 0.00 | 0.46 |
| 3 | 3.63 | 0.49 | 3.03 | 0.18 | 3.00 | 0.00 | -0.63 | -0.03 | 0.00 | 0.47 |
| 3 | 3.73 | 0.45 | 3.13 | 0.35 | 2.97 | 0.18 | -0.73 | -0.13 | 0.03 | 0.48 |
| 3 | 3.90 | 0.31 | 3.43 | 0.50 | 3.00 | 0.00 | -0.90 | -0.43 | 0.00 | 0.53 |
| 3 | 3.73 | 0.45 | 3.07 | 0.25 | 3.00 | 0.00 | -0.73 | -0.07 | 0.00 | 0.55 |
| 3 | 3.93 | 0.25 | 3.43 | 0.50 | 3.00 | 0.00 | -0.93 | -0.43 | 0.00 | 0.57 |
| 3 | 3.97 | 0.18 | 3.50 | 0.51 | 3.00 | 0.00 | -0.97 | -0.50 | 0.00 | 0.61 |
| 3 | 3.87 | 0.35 | 3.17 | 0.38 | 3.00 | 0.00 | -0.87 | -0.17 | 0.00 | 0.61 |
| **3** | **3.90** | **0.31** | **3.17** | **0.38** | **3.00** | **0.00** | **-0.90** | **-0.17** | **0.00** | **0.66** |
| **3** | **3.90** | **0.31** | **3.13** | **0.35** | **3.00** | **0.00** | **-0.90** | **-0.13** | **0.00** | **0.69** |
| **4** | **3.17** | **0.38** | **3.10** | **0.31** | **2.97** | **0.18** | **0.83** | **0.90** | **1.03** | **0.04** |
| **4** | **3.47** | **0.51** | **3.27** | **0.45** | **2.97** | **0.18** | **0.53** | **0.73** | **1.03** | **0.18** |
| 4 | 3.50 | 0.51 | 3.20 | 0.41 | 3.00 | 0.00 | 0.50 | 0.80 | 1.00 | 0.21 |
| 4 | 3.47 | 0.51 | 3.00 | 0.26 | 3.00 | 0.00 | 0.53 | 1.00 | 1.00 | 0.30 |
| 4 | 3.57 | 0.50 | 3.13 | 0.35 | 3.00 | 0.00 | 0.43 | 0.87 | 1.00 | 0.30 |
| 4 | 3.60 | 0.50 | 3.17 | 0.38 | 3.00 | 0.00 | 0.40 | 0.83 | 1.00 | 0.31 |
| 4 | 3.50 | 0.51 | 3.03 | 0.18 | 3.00 | 0.00 | 0.50 | 0.97 | 1.00 | 0.33 |
| 4 | 3.83 | 0.38 | 3.47 | 0.51 | 3.00 | 0.00 | 0.17 | 0.53 | 1.00 | 0.46 |
| 4 | 3.90 | 0.31 | 3.53 | 0.51 | 3.00 | 0.00 | 0.10 | 0.47 | 1.00 | 0.53 |
| 4 | 3.90 | 0.31 | 3.30 | 0.47 | 3.00 | 0.00 | 0.10 | 0.70 | 1.00 | 0.57 |
| 4 | 3.93 | 0.25 | 3.40 | 0.50 | 3.00 | 0.00 | 0.07 | 0.60 | 1.00 | 0.58 |
| 4 | 4.00 | 0.00 | 3.60 | 0.50 | 3.00 | 0.00 | 0.00 | 0.40 | 1.00 | 0.67 |
| **4** | **3.93** | **0.25** | **3.80** | **0.48** | **3.00** | **0.00** | **0.07** | **0.20** | **1.00** | **0.68** |
| **4** | **3.93** | **0.25** | **3.03** | **0.18** | **3.00** | **0.00** | **0.07** | **0.97** | **1.00** | **0.85** |